\documentclass[twocolumn,superscriptaddress,letter,prb]{revtex4}%
\usepackage{amssymb}
\usepackage{color}
\usepackage{graphicx}
\usepackage{dcolumn}
\usepackage{bm}
\usepackage[header,title,page,titletoc]{appendix}
\usepackage{amsmath}
\usepackage{amsfonts}%
\providecommand{\U}[1]{\protect \rule{.1in}{.1in}}
\begin{document}
\title{Topological Superfluid in P-band Optical Lattice}
\author{Ya-Jie Wu}
\affiliation{Department of Physics, Beijing Normal University, Beijing 100875, China}
\author{Jing He}
\affiliation{Department of Physics, Beijing Normal University, Beijing 100875, China}
\author{Chun-Li Zang}
\affiliation{Department of Physics, Beijing Normal University, Beijing 100875, China}
\author{Su-Peng Kou}
\thanks{Corresponding author}
\email{spkou@bnu.edu.cn}
\affiliation{Department of Physics, Beijing Normal University, Beijing 100875, China}

\begin{abstract}
In this paper by studying p-band fermionic system with nearest neighbor
attractive interaction we find translation symmetry protected $%
%TCIMACRO{\U{2124} }%
%BeginExpansion
\mathbb{Z}
%EndExpansion
_{2}$ topological superfluid (TSF) that is characterized by a special fermion
parity pattern at high symmetry points in momentum space $\mathbf{k}=(0,0)$,
$(0,\pi)$, $(\pi,0)$, $(\pi,\pi)$. Such $%
%TCIMACRO{\U{2124} }%
%BeginExpansion
\mathbb{Z}
%EndExpansion
_{2}$ TSF supports the robust Majorana edge modes and a new type of low energy
excitation - (supersymmetric) $%
%TCIMACRO{\U{2124} }%
%BeginExpansion
\mathbb{Z}
%EndExpansion
_{2}$ link-excitation. In the end we address a possible realization of such
interacting p-band fermions with $%
%TCIMACRO{\U{2124} }%
%BeginExpansion
\mathbb{Z}
%EndExpansion
_{2}$ TSF.


\end{abstract}
\maketitle

Superconductivity/superfluidity is a paired state in
many-body\ charged/neutral fermion systems introduced by Bardeen, Cooper and
Schrieffer (BCS). In BCS's theory, the glue for the Cooper pair is phonon -
the collective mode of atoms in solids. To describe such a quantum ordered
state, a local order parameter $\Delta_{\mathbf{k}}=\left \langle \hat
{c}_{\mathbf{k}}\hat{c}_{-\mathbf{k}}\right \rangle $ is introduced that
differs in different superconducting (SC) states (for example, s-wave, p-wave,
d-wave, etc.). Recently, people found that SC states with the same local order
parameter may have different topological properties, that leads to the concept
of "topological superconductivity (TSC)"\cite{vol}. According to the
characterization of "ten-fold way" of random matrix\cite{ry}, there are three
types of TSCs in two dimensions : D-type chiral TSC without time reversal
symmetry ($p_{x}\pm \mathrm{i}p_{y}$-wave\cite{read}, s-wave with strong
spin-orbital coupling\cite{so}, ...), C-type chiral TSC without time reversal
symmetry ($d_{x}\pm \mathrm{i}d_{y}$-wave \cite{1}) and DIII-type $%
%TCIMACRO{\U{2124} }%
%BeginExpansion
\mathbb{Z}
%EndExpansion
_{2}$ TSC with time reversal symmetry ($\left(  p_{x}+\mathrm{i}p_{y}\right)
\times \left(  p_{x}-\mathrm{i}p_{y}\right)  $-wave SC\cite{roy}).

Recently, rapid advances in trapping and manipulating ultrocold atoms in
optical lattice\ have made it possible to simulate topological superfluid
(TSF) - a kind of "SC" in a neutral fermion system. For the paired states in
optical lattices, the glues cannot be the phonons. Instead, people have more
degrees of freedom to tune the interaction: the tunable direct attractive
interaction by Feshbach resonance (see reviews in \cite{fesh}), the effective
attractive interaction of composite fermions in Bose-Fermion mixture from
background Bosons\cite{mix1}, the anisotropic dipole-dipole
interaction\cite{dipole}. Because the non-Abelian anyon of D-type chiral TSF
may be designed as a decoherence-free qubit and plays an important role in the
realization of fault-tolerant topological quantum computation\cite{ki2,free},
people pay more attention on its realization. For instance, for spinless
(s-band) fermions in an optical lattice with the nearest neighbor attractive
interaction, the ground state may be chiral $p_{x}\pm \mathrm{i}p_{y}$-wave
TSC\cite{spinless,lo}.

In optical lattices, besides the tunable interactions, people may have an
additional degree of freedom to obtain the TSC by considering p-band
fermions\cite{po}. In this paper we studied (spinless) p-band fermions with
nearest neighbor attractive interaction on a square optical lattice and found
that the ground state may be a $p_{x}\times p_{y}$-wave SF - the $p_{x}$-band
fermions are paired into $p_{x}$-wave SF state and the $p_{y}$-band fermions
are paired into $p_{y}$-wave SF state. Such a $p_{x}\times p_{y}$-wave paired
state is translation symmetry protected $%
%TCIMACRO{\U{2124} }%
%BeginExpansion
\mathbb{Z}
%EndExpansion
_{2}$\emph{\ }TSF against arbitrary translation invariant perturbations. In
addition, we found that this $%
%TCIMACRO{\U{2124} }%
%BeginExpansion
\mathbb{Z}
%EndExpansion
_{2}$ TSF supports the robust Majorana edge modes and there exists a new type
of low energy excitations - (supersymmetric) $%
%TCIMACRO{\U{2124} }%
%BeginExpansion
\mathbb{Z}
%EndExpansion
_{2}$ link-excitations. Let us explain why there exist $%
%TCIMACRO{\U{2124} }%
%BeginExpansion
\mathbb{Z}
%EndExpansion
_{2}$\ TSF beyond "ten-fold way" : if one enforces translation invariant on
the system, there may exist sixteen symmetry protected TSCs/TSFs on a two
dimensional (2D) lattice that are characterized by a special fermion parity
pattern at four high symmetry points ($\mathbf{k}=(0,0)$, $(0,\pi)$, $(\pi
,0)$, $(\pi,\pi)$)\cite{kou}.

\textit{Mean field phase diagram and }$%
%TCIMACRO{\U{2124} }%
%BeginExpansion
\mathbb{Z}
%EndExpansion
_{2}$\textit{\ TSF}: In this paper our starting point is the following
effective model of spinless p-band fermions with on-site and nearest neighbor
attractive interactions on square lattice
\begin{align}
H_{\mathrm{eff}}  &  =t\sum \limits_{i}\hat{c}_{i\uparrow}^{\dagger}\hat
{c}_{i+e_{x}\uparrow}+t\sum \limits_{i}\hat{c}_{i\downarrow}^{\dagger}\hat
{c}_{i+e_{y}\downarrow}-U\sum \limits_{i}\hat{n}_{i\uparrow}\hat{n}%
_{i\downarrow}\nonumber \\
&  -U_{0}\sum \limits_{\left \langle i,j\right \rangle }\hat{n}_{i}\hat{n}%
_{j}-\mu \sum \limits_{i\sigma}\hat{c}_{i\sigma}^{\dagger}\hat{c}_{i\sigma}
\label{1}%
\end{align}
where $p_{x}$-band fermions only hop along x-direction and $p_{y}$-band
fermions only hop along y-direction. $U$ and $U_{0}$ are the on-site
interaction strength and the nearest neighbor interaction strength,
respectively ($U>0,$ $U_{0}>0$). We denote orbitals index by pseudo-spin
$\sigma=(p_{x}$, $p_{y})\equiv(\uparrow$, $\downarrow)$ and $\left \langle
{i,j}\right \rangle $ represents two sites on a nearest-neighbor link. In this
paper we only consider the half filling case with $\mu=0$.

The model in Eq.(\ref{1}) is unstable against superfluid (SF) orders that are
described by $\triangle_{s}$ and $\triangle_{p}$ for s-wave and p-wave order
parameters, where $\Delta_{s}=\left \langle \hat{c}_{i\uparrow}\hat
{c}_{i\downarrow}\right \rangle $, $\Delta_{p}=\left \langle \hat{c}_{i\uparrow
}\hat{c}_{i+e_{x}\uparrow}\right \rangle =\left \langle \hat{c}_{i\downarrow
}\hat{c}_{i+e_{y}\downarrow}\right \rangle $, respectively. Then the effective
Hamiltonian becomes
\begin{align}
H_{\mathrm{eff}}  &  =t\sum \limits_{i}\hat{c}_{i\uparrow}^{\dagger}\hat
{c}_{i+e_{x}\uparrow}-U_{0}\Delta_{p}\sum \limits_{i}\hat{c}_{i\uparrow
}^{\dagger}\hat{c}_{i+e_{x}\uparrow}^{\dagger}-U\Delta_{s}\sum \limits_{i}%
\hat{c}_{i\uparrow}\hat{c}_{i\downarrow}\nonumber \\
&  +t\sum \limits_{i}\hat{c}_{i\downarrow}^{\dagger}\hat{c}_{i+e_{y}\downarrow
}-U_{0}\Delta_{p}\sum \limits_{i}\hat{c}_{i\downarrow}^{\dagger}\hat
{c}_{i+e_{y}\downarrow}^{\dagger}+h.c..
\end{align}
One can see that $p_{x}$-band fermions only pair along x-direction,
$\left \langle \hat{c}_{i\uparrow}\hat{c}_{i+e_{x}\uparrow}\right \rangle \neq
0$, $\left \langle \hat{c}_{i\uparrow}\hat{c}_{i+e_{y}\uparrow}\right \rangle
=0$ and $p_{y}$-band fermions only pair along y-direction, $\left \langle
\hat{c}_{i\downarrow}\hat{c}_{i+e_{y}\downarrow}\right \rangle \neq0$,
$\left \langle \hat{c}_{i\downarrow}\hat{c}_{i+e_{x}\downarrow}\right \rangle
=0$. For this reason we call the p-wave SF as $p_{x}\times p_{y}$-wave state.
In the SF state, the total fermion number $N_{F}=\sum \nolimits_{i}n_{i}$ is
conserved only mod $2$. Thus the fermion parity $P_{F}=(-1)^{N_{F}}$ is good
quantum number.

\begin{figure}[ptb]
\includegraphics[width=0.45\textwidth]{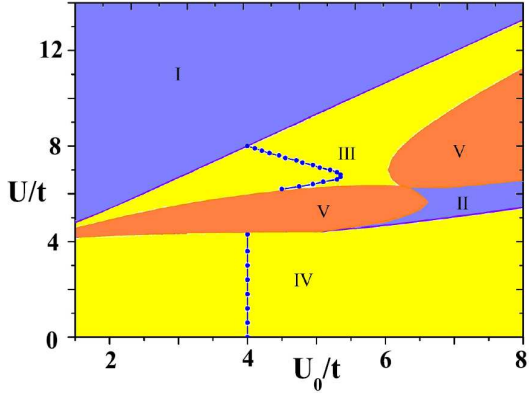}\caption{Phase diagram: region
I and region II are trivial SC states with $\Delta_{s}\neq0,$ $\Delta_{p}=0$
and $\Delta_{s}\neq0,$ $\Delta_{p}\neq0$, respectively; Yellow regions (region
III and region IV) are $\mathbb{Z}_{2}$ TSFs with $\Delta_{s}\neq0,$
$\Delta_{p}\neq0$ and $\Delta_{s}=0,$ $\Delta_{p}\neq0$, respectively; region
V is a gapless state that we are not interest in. The blue lines with dots are
the super-$\mathbb{Z}_{2}$ TSFs.}%
\end{figure}

After solving self-consistent equations by minimizing the ground state energy,
we arrive at the mean-field phase diagram shown in Fig.1. In Fig.1, there are
five regions : I, II, III, IV, V. To classify the topological properties of
the ground states in the five regions, we calculate fermionic parities at high
symmetry points by defining four $%
%TCIMACRO{\U{2124} }%
%BeginExpansion
\mathbb{Z}
%EndExpansion
_{2}$ topological invariables\cite{kou}: for $\mathbf{k=}(0,0),$ $(\pi,\pi),$
we have a trivial result as $\mathcal{\zeta}_{\mathbf{k}=(0,0)}=\mathcal{\zeta
}_{\mathbf{k}=(\pi,\pi)}=0;$ for $\mathbf{k=}(0,\pi),$ $(\pi,0),$ we have
\begin{equation}
\mathcal{\zeta}_{\mathbf{k}=(0,\pi)}=\mathcal{\zeta}_{\mathbf{k}=(\pi
,0)}=1-\Theta \left(  \mu^{2}+U^{2}\Delta_{s}^{2}-4t^{2}\right)
\end{equation}
where $\Theta(x)=1$ if $x>0$ and $\Theta(x)=0$ if $x<0$. Thus we identify two
distinct SF states : SF with trivial topological indices $\mathcal{\zeta
}_{\mathbf{k}=(0,\pi)}=\mathcal{\zeta}_{\mathbf{k}=(\pi,0)}=0$ in region I
($\Delta_{s}\neq0,$ $\Delta_{p}=0$) and region II ($\Delta_{s}\neq0,$
$\Delta_{p}\neq0$) and $%
%TCIMACRO{\U{2124} }%
%BeginExpansion
\mathbb{Z}
%EndExpansion
_{2}$\textit{\ }TSFs with nontrivial topological indices $\mathcal{\zeta
}_{\mathbf{k}=(0,\pi)}=\mathcal{\zeta}_{\mathbf{k}=(\pi,0)}=1$ in region III
($\Delta_{s}\neq0,$ $\Delta_{p}\neq0$) and IV ($\Delta_{s}=0,$ $\Delta_{p}%
\neq0$). From topological indices of $%
%TCIMACRO{\U{2124} }%
%BeginExpansion
\mathbb{Z}
%EndExpansion
_{2}$\textit{\ }TSF ($0110$),\cite{kou} we found a special fermion parity
pattern: even fermion parity at $\mathbf{k}=(0,0)$ and $\mathbf{k}=(\pi,\pi)$
and odd fermion parity at $\mathbf{k}=(0,\pi)$ and $\mathbf{k}=(\pi,0)$. This
is the reason why we call it $%
%TCIMACRO{\U{2124} }%
%BeginExpansion
\mathbb{Z}
%EndExpansion
_{2}$ TSF. The SF states in region V ($\Delta_{s}\neq0,$ $\Delta_{p}\neq0$) is
gapless which we are not interest in. In the following parts, we will focus on
$%
%TCIMACRO{\U{2124} }%
%BeginExpansion
\mathbb{Z}
%EndExpansion
_{2}$\textit{\ }TSFs in region III and region IV.

\begin{figure}[ptb]
\includegraphics[width=0.48\textwidth]{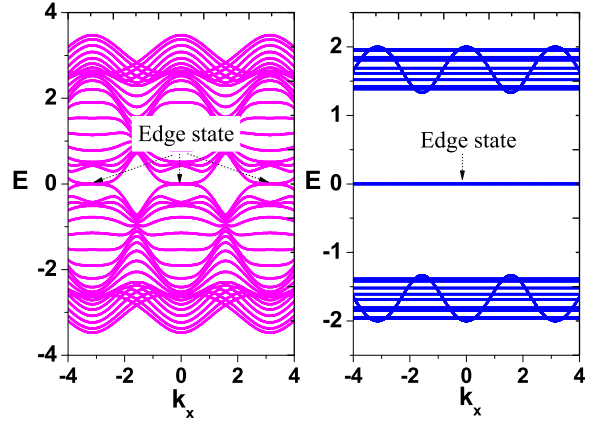}\caption{Left figure: Edge
states of TSF in region III which have a symmetry protected zero energy mode
at $k_{x}=0,\pi$; Right figure: Edge states of TSF in region IV which have a
flat band.}%
\end{figure}

\textit{Majorana edge state}: In this part, we study the edge states of $%
%TCIMACRO{\U{2124} }%
%BeginExpansion
\mathbb{Z}
%EndExpansion
_{2}$\textit{\ }TSFs. Fig.2 illustrates the edge states in region III and IV.
In the region IV ($\Delta_{s}=0,$ $\Delta_{p}\neq0$), $%
%TCIMACRO{\U{2124} }%
%BeginExpansion
\mathbb{Z}
%EndExpansion
_{2}$\textit{\ }TSF behaviors likes two coupled 1D p-wave SCs, one along
x-direction, and the other along y-direction. Because each 1D p-wave SCs with
effective Hamiltonian $H_{\mathrm{eff}}=t\sum \nolimits_{i}\hat{c}%
_{\uparrow/\downarrow,i}^{\dagger}\hat{c}_{\uparrow/\downarrow,i+e_{x/y}%
}-\Delta_{p}U_{0}\hat{c}_{\uparrow/\downarrow,i}\hat{c}_{\uparrow
/\downarrow,i+e_{x/y}}+h.c.$ is really a Majorana chain, we get zero energy
states at the boundary\cite{Kitaev}. Indeed, the zero energy states can be
mapped to a Majorana fermion system with a flat band exactly (See right figure
in Fig.2).\ In the region of III ($\Delta_{s}\neq0,$ $\Delta_{p}\neq0$), the
Majorana fermion system is still stable but has dispersion\cite{kou2} (See
left figure in Fig.2).

On the contrary, in the region I and II, there are no edge states at all.

$%
%TCIMACRO{\U{2124} }%
%BeginExpansion
\mathbb{Z}
%EndExpansion
_{2}$\textit{\ link-excitation: }Next, we study the low energy excitations in
the $%
%TCIMACRO{\U{2124} }%
%BeginExpansion
\mathbb{Z}
%EndExpansion
_{2}$ TSF states in region III and IV. One excitation is the Bogoliubov
quasi-particle which has an energy gap $E_{qp}$ in $%
%TCIMACRO{\U{2124} }%
%BeginExpansion
\mathbb{Z}
%EndExpansion
_{2}$\textit{\ }TSFs (See blue line in Fig.3). Another excitation is the
collective mode - the Goldstone mode due to spontaneous global \textrm{U(1)}
symmetry breaking. In this part we emphasis that there may exist another type
of low energy excitations for TSF on a square lattice -- the $%
%TCIMACRO{\U{2124} }%
%BeginExpansion
\mathbb{Z}
%EndExpansion
_{2}$ link-excitation.

A $%
%TCIMACRO{\U{2124} }%
%BeginExpansion
\mathbb{Z}
%EndExpansion
_{2}$ link-excitation is defined as the sign-switching of SC order parameter
on a link position $\mathbf{I}=(I,J)=\left(  i_{0},i_{0}+e_{x}\mathbf{/}%
e_{y}\right)  $ as $\Delta_{p}\rightarrow \{%
\begin{array}
[c]{c}%
-\Delta_{p},\text{ }i=i_{0}\\
\Delta_{p}\text{, }i\neq i_{0}%
\end{array}
$. From this definition, one can see that its particle number is conserved mod
$2.$ This is the reason why we call it $%
%TCIMACRO{\U{2124} }%
%BeginExpansion
\mathbb{Z}
%EndExpansion
_{2}$ link-excitation. By exact diagonalization calculations, we obtain the
excited energy $E_{link}$ for a single $%
%TCIMACRO{\U{2124} }%
%BeginExpansion
\mathbb{Z}
%EndExpansion
_{2}$ link-excitation in Fig.3 (See magenta line). From it one can see that
the $%
%TCIMACRO{\U{2124} }%
%BeginExpansion
\mathbb{Z}
%EndExpansion
_{2}$ link-excitation even have lower energy than quasi-particle excitation
and becomes relevant to low energy physics.

\begin{figure}[ptb]
\includegraphics[width=0.48\textwidth]{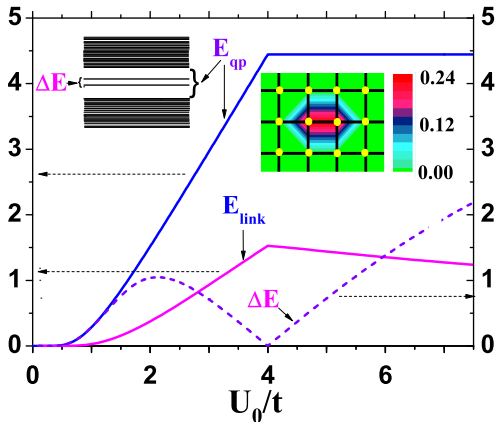}\caption{Three energy scales
in the TSF states : $E_{qp}$, $E_{link}$, $\Delta E$ (The scale is t). The
left inset is the scheme of the energy spectrum of quasi-particles, of which
$E_{qp}$ is the energy gap of Bogoliubov quasi-particle and $\Delta E$ is the
energy splitting of the bound state of $\mathbb{Z} _{2}$ link-excitation. The
right inset figure shows the bound state on a $\mathbb{Z} _{2}$
link-excitation with $U/t=0.3$, $U_{0}/t=3.25$. $E_{link}$ is the excited
energy of a $\mathbb{Z} _{2}$ link-excitation.}%
\end{figure}

We point out that an important property of $%
%TCIMACRO{\U{2124} }%
%BeginExpansion
\mathbb{Z}
%EndExpansion
_{2}$ link-excitation is the existence of a bound state (zero mode) on it for
the $%
%TCIMACRO{\U{2124} }%
%BeginExpansion
\mathbb{Z}
%EndExpansion
_{2}$ TSFs states both in region III and IV. The inset of Fig.3 shows the
bound state on a $%
%TCIMACRO{\U{2124} }%
%BeginExpansion
\mathbb{Z}
%EndExpansion
_{2}$ link-excitation in the region IV. Then we define that $E_{+}$ is the
energy for the empty bound state (or the even parity state) and $E_{-}$ is the
energy for the occupied bound state (or odd parity state). In the region IV,
when $U_{0}\Delta_{p}\neq t,$ there may exist energy splitting between $E_{+}$
and $E_{-}$ due to quantum tunneling effect, $\Delta E=E_{+}-E_{-}.$ By exact
diagonalization calculations, we derive $\Delta E$ in Fig.3 (See dash violet
line). The occupied bound state with odd fermion parity has lower energy than
the empty bound state with even fermion parity, $\Delta E>0$ (we have set the
chemical potential to be zero). Now the $%
%TCIMACRO{\U{2124} }%
%BeginExpansion
\mathbb{Z}
%EndExpansion
_{2}$ link-excitation can be regarded as a fermion when the energy scale is
smaller than $\Delta E.$

Particularly, in the region IV, when $U_{0}\Delta_{p}=t$, the energy splitting
of the two bound states becomes zero $\Delta E=0$. For this case, the energy
dispersion of quasi-particles becomes a constant or the Bogoliubov
quasi-particles have a flat band, that corresponds to a quasi-particle with
divergent mass. Thus the quantum tunneling effect by exchanging
quasi-particles is frozen and the energy is degenerate for an occupied zero
mode and an empty zero mode. Now the $%
%TCIMACRO{\U{2124} }%
%BeginExpansion
\mathbb{Z}
%EndExpansion
_{2}$ link-excitation of even parity bound state is a Boson and that of odd
parity state is a Fermion, that is really \emph{emergent supersymmetry}. And
we call this TSF with supersymmetric $%
%TCIMACRO{\U{2124} }%
%BeginExpansion
\mathbb{Z}
%EndExpansion
_{2}$\ link-excitation to be \emph{super-}$%
%TCIMACRO{\U{2124} }%
%BeginExpansion
\mathbb{Z}
%EndExpansion
_{2}$\emph{\ TSF}.

We also study the $%
%TCIMACRO{\U{2124} }%
%BeginExpansion
\mathbb{Z}
%EndExpansion
_{2}$ link-excitations in region III and find that their properties are the
same as those in region IV (see the dotted line in region III of Fig.1 that
corresponds to the super-$%
%TCIMACRO{\U{2124} }%
%BeginExpansion
\mathbb{Z}
%EndExpansion
_{2}$ TSF).

$%
%TCIMACRO{\U{2124} }%
%BeginExpansion
\mathbb{Z}
%EndExpansion
_{2}$\textit{\ vortex}: After recognizing the properties of $%
%TCIMACRO{\U{2124} }%
%BeginExpansion
\mathbb{Z}
%EndExpansion
_{2}$ link-excitation, we turn to study $%
%TCIMACRO{\U{2124} }%
%BeginExpansion
\mathbb{Z}
%EndExpansion
_{2}$ vortex. For a SF on square lattice, $%
%TCIMACRO{\U{2124} }%
%BeginExpansion
\mathbb{Z}
%EndExpansion
_{2}$ vortex is really a $\pi$-flux on a plaquette which is confined at zero
temperature and cannot be real excitation. On one hand, the $%
%TCIMACRO{\U{2124} }%
%BeginExpansion
\mathbb{Z}
%EndExpansion
_{2}$ link-excitations can be regarded as bound states of a pair of $%
%TCIMACRO{\U{2124} }%
%BeginExpansion
\mathbb{Z}
%EndExpansion
_{2}$ vortices\textit{\ }on nearest neighbor plaquettes ($A$ and $B$) (See
Fig.4). On the other hand, the $%
%TCIMACRO{\U{2124} }%
%BeginExpansion
\mathbb{Z}
%EndExpansion
_{2}$ vortex can be regarded as the end of non-local string of $%
%TCIMACRO{\U{2124} }%
%BeginExpansion
\mathbb{Z}
%EndExpansion
_{2}$ link-excitations and is generated by a string of link-excitations.

When $%
%TCIMACRO{\U{2124} }%
%BeginExpansion
\mathbb{Z}
%EndExpansion
_{2}$ link-excitation is a fermion ($U_{0}\Delta_{p}\neq t$), due to fermion
parity constraint, $%
%TCIMACRO{\U{2124} }%
%BeginExpansion
\mathbb{Z}
%EndExpansion
_{2}$ vortex cannot hop from a plaquette to nearest neighbor plaquette by
creating an additional fermionic $%
%TCIMACRO{\U{2124} }%
%BeginExpansion
\mathbb{Z}
%EndExpansion
_{2}$ link-excitation which violates the fermion parity conservation. Thus
there must exist two types of $%
%TCIMACRO{\U{2124} }%
%BeginExpansion
\mathbb{Z}
%EndExpansion
_{2}$ vortices: $e$-type $%
%TCIMACRO{\U{2124} }%
%BeginExpansion
\mathbb{Z}
%EndExpansion
_{2}$ vortex on $A$ sub-plaquette\ and the $m$-type $%
%TCIMACRO{\U{2124} }%
%BeginExpansion
\mathbb{Z}
%EndExpansion
_{2}$ vortex on $B$ sub-plaquette. Each type of $%
%TCIMACRO{\U{2124} }%
%BeginExpansion
\mathbb{Z}
%EndExpansion
_{2}$ vortex lies on each sub-plaquette. In addition, there is mutual-semion
statistics between $e$-type $%
%TCIMACRO{\U{2124} }%
%BeginExpansion
\mathbb{Z}
%EndExpansion
_{2}$ vortex and $m$-type $%
%TCIMACRO{\U{2124} }%
%BeginExpansion
\mathbb{Z}
%EndExpansion
_{2}$ vortex. And an $e$-type $%
%TCIMACRO{\U{2124} }%
%BeginExpansion
\mathbb{Z}
%EndExpansion
_{2}$ vortex and an $m$-type $%
%TCIMACRO{\U{2124} }%
%BeginExpansion
\mathbb{Z}
%EndExpansion
_{2}$ vortex annihilate with each other into a fermionic $%
%TCIMACRO{\U{2124} }%
%BeginExpansion
\mathbb{Z}
%EndExpansion
_{2}$ link-excitation. Now the system has $4$ superselection sectors: $1$ (the
vacuum), $e$ ($e$-type $%
%TCIMACRO{\U{2124} }%
%BeginExpansion
\mathbb{Z}
%EndExpansion
_{2}$ vortex), $m$ ($m$-type $%
%TCIMACRO{\U{2124} }%
%BeginExpansion
\mathbb{Z}
%EndExpansion
_{2}$ vortex), $\psi$ (fermionic $%
%TCIMACRO{\U{2124} }%
%BeginExpansion
\mathbb{Z}
%EndExpansion
_{2}$ link-excitation) and the fusion rule is $e\times e=m\times m=\psi
\times \psi=1,$ $e\times m=\psi,\quad e\times \psi=m,\quad m\times \psi=e.$ The
fusion rule is the same as that of the $%
%TCIMACRO{\U{2124} }%
%BeginExpansion
\mathbb{Z}
%EndExpansion
_{2}$ topological order in toric code model and Wen's plaquette
model\cite{ki1,wen1}.

\begin{figure}[ptb]
\includegraphics[width=0.52\textwidth]{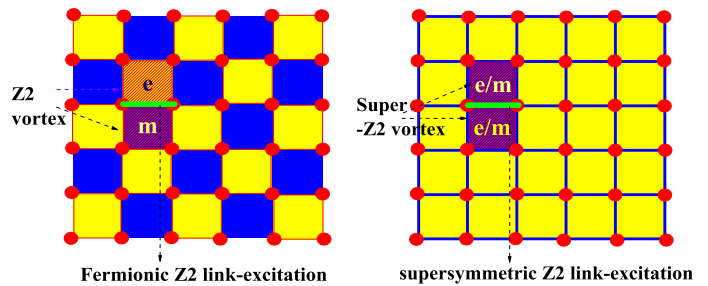}\caption{Left figure:
Illustration of the relationship between fermionic $\mathbb{Z} _{2}$
link-excitation (the green link) and $e$-type $\mathbb{Z} _{2}$ vortex,
$m$-type $\mathbb{Z} _{2}$ vortex (the shadow plaquettes) of $\mathbb{Z} _{2}%
$\ TSF. There are two sub-plaquettes, one for $e$-type $\mathbb{Z} _{2}$
vortex, the other for $m$-type $\mathbb{Z} _{2}$ vortex; Right figure:
Illustration of the relationship between supersymmetric $\mathbb{Z} _{2}$
link-excitation (the green link) and two-component super-$\mathbb{Z} _{2}$
vortex (the shadow plaquettes) of super-$\mathbb{Z} _{2}$\ TSF.}%
\end{figure}

For the super-$%
%TCIMACRO{\U{2124} }%
%BeginExpansion
\mathbb{Z}
%EndExpansion
_{2}$\ TSF, we find that $%
%TCIMACRO{\U{2124} }%
%BeginExpansion
\mathbb{Z}
%EndExpansion
_{2}$ vortex has internal degree of freedom - it can be either $e$-type $%
%TCIMACRO{\U{2124} }%
%BeginExpansion
\mathbb{Z}
%EndExpansion
_{2}$ vortex or $m$-type $%
%TCIMACRO{\U{2124} }%
%BeginExpansion
\mathbb{Z}
%EndExpansion
_{2}$ vortex. That is each $%
%TCIMACRO{\U{2124} }%
%BeginExpansion
\mathbb{Z}
%EndExpansion
_{2}$ vortex has two components as $\mathbf{\sigma}=(%
\begin{array}
[c]{c}%
e\\
m
\end{array}
)$. When same types of $%
%TCIMACRO{\U{2124} }%
%BeginExpansion
\mathbb{Z}
%EndExpansion
_{2}$-vortices ($ee,$ $mm$) annihilate each other, we have a bosonic $%
%TCIMACRO{\U{2124} }%
%BeginExpansion
\mathbb{Z}
%EndExpansion
_{2}$ link-excitation which belongs to the vacuum sector; When different types
of $%
%TCIMACRO{\U{2124} }%
%BeginExpansion
\mathbb{Z}
%EndExpansion
_{2}$-vortices ($em$) annihilate each other, we have a fermionic $%
%TCIMACRO{\U{2124} }%
%BeginExpansion
\mathbb{Z}
%EndExpansion
_{2}$ link-excitation. For this reason we call it\emph{\ }super-$%
%TCIMACRO{\U{2124} }%
%BeginExpansion
\mathbb{Z}
%EndExpansion
_{2}$-vortex. See the schemes in Fig.4. The nonAbelian representation of the
fusion rule for the supersymmetric case is $\mathbf{\sigma}\times
\mathbf{\sigma}=1+\psi$ where $\mathbf{\sigma}$ is a matrix. This equation is
much similar to that of the nonAbelian TSC\cite{mr,iva} where the vortex
section $\sigma$ is not a matrix. Just for this reason, super-$%
%TCIMACRO{\U{2124} }%
%BeginExpansion
\mathbb{Z}
%EndExpansion
_{2}$ TSF may be another possible candidate for the fault-tolerant topological
quantum computation.

\textit{Physics realization:} In general, to realize such exotic $%
%TCIMACRO{\U{2124} }%
%BeginExpansion
\mathbb{Z}
%EndExpansion
_{2}$ TSF, we need to design a spinless p-orbital fermionic model with
significant nearest neighbor attractive interaction. Unfortunately, for p-band
fermions the direct nearest neighbor attracting interaction is too small to be
considered. A candidate is an interacting mixture of ultracold bosons on
s-band and spinless fermions on p-band in 2D square optical lattice\cite{mix1}%
, of which the effective Hamiltonian is$\ H=\sum \nolimits_{\delta
=s,p}(H_{\delta}-\mu_{\delta}\hat{N}_{\delta})-U_{\mathrm{sp}}\sum
\nolimits_{i}\hat{n}_{s,i}\hat{n}_{p,i}$ where $H_{s}$ and $H_{p}$ describes
the hopping terms of s-obital bosons and p-obital fermions: $H_{s}=-t_{s}%
\sum \nolimits_{\left \langle i,j\right \rangle }\hat{b}_{s,i}^{\dagger}\hat
{b}_{s,j}+\frac{U_{s}}{2}\sum \nolimits_{i}\hat{n}_{s,i}(\hat{n}_{s,i}-1)$ and
$H_{p}=t_{\parallel}\sum \nolimits_{i}\hat{f}_{i\uparrow}^{\dagger}\hat
{f}_{i+e_{x}\uparrow}+t_{\parallel}\sum \nolimits_{i}\hat{f}%
_{i\mathbf{\downarrow,}}^{\dagger}\hat{f}_{i+e_{y}\downarrow}-t_{\bot}%
\sum \nolimits_{i}\hat{f}_{i\mathbf{\downarrow,}}^{\dagger}\hat{f}%
_{i+e_{x}\downarrow}-t_{\bot}\sum \nolimits_{i}\hat{f}_{i\mathbf{\uparrow,}%
}^{\dagger}\hat{f}_{i+e_{y}\uparrow}$, respectively. $t_{s}$ is the hopping
parameter for s-band bosons and $t_{\parallel}$/$t_{\bot}$ is hopping
parameter for p-band fermions on neighboring sites parallel/perpendicular to
the bond direction. Pseudo-spin $(\uparrow$, $\downarrow)$ denotes orbitals
index $(p_{x}$, $p_{y}).$ Due to $t_{\bot}\ll t_{\parallel},$ we set $t_{\bot
}$ to be zero. $U_{\mathrm{sp}}>0$ denotes the attractive interaction strength
between s-band bosons and p-band fermions and $U_{s}>0$ denotes the repulsive
interaction strength between s-band bosons. $\mu_{\delta}$ is the chemical
potential and $\hat{N}_{\delta}=\sum \limits_{i}\hat{n}_{\delta,i}$ is the
particle number operator where $\hat{n}_{s,i}=\hat{b}_{s,i}^{\dagger}\hat
{b}_{s,j}$ and $\hat{n}_{p,i}=\hat{f}_{i\uparrow}^{\dagger}\hat{f}_{i\uparrow
}+\hat{f}_{i\mathbf{\downarrow,}}^{\dagger}\hat{f}_{i\downarrow}$. Here, we
consider the Mott insulator phase of s-band bosons with unit filling and a
half filling case of p-band fermions.

By tuning the interaction between s-orbital bosons and p-orbital fermions for
the case of $U_{s}>U_{\mathrm{sp}}>0$, an s-orbital boson and a p-orbital
fermion may bind into a composite p-orbital fermion with both effective
on-site and effective nearest neighbor attractive interaction (See detailed
calculations in Ref.\cite{mix1}). In this case, the corresponding parameters
in Eq.[1] are $t\rightarrow2t_{s}t_{\parallel}/U_{\mathrm{sp}}$, $U\rightarrow
U_{s}-4U_{\mathrm{sp}},$ $U_{0}\rightarrow-2t_{s}^{2}/U_{s}-t_{s}%
^{2}/U_{\mathrm{sp}}-t_{\parallel}^{2}/U_{\mathrm{sp}},$
respectively\cite{note}. Here $\hat{c}_{i}$ is the operator of the composite
fermion as $\hat{b}_{i}\hat{f}_{i}.$ From the fact $U_{0}<0$, there always
exists effective nearest neighbor attractive interaction that will leads to
$p_{x}\times p_{y}$-wave pairing between composite fermion of the same
p-orbital. If $4U_{\mathrm{sp}}>U_{s}>U_{\mathrm{sp}},$ we have an additional
on-site attractive interaction that will leads to s-wave pairing between
composite fermion of the opposite p-orbital. When the effective nearest
neighbor attracting interaction dominates, the ground state is just the $%
%TCIMACRO{\U{2124} }%
%BeginExpansion
\mathbb{Z}
%EndExpansion
_{2}$ TSF.

\textit{Conclusion : }In this paper we studied the interacting p-band fermions
on square lattice and found the ground state may be translation symmetry
protected $%
%TCIMACRO{\U{2124} }%
%BeginExpansion
\mathbb{Z}
%EndExpansion
_{2}$ topological superfluid beyond the characterization of "ten-fold way".
Such $%
%TCIMACRO{\U{2124} }%
%BeginExpansion
\mathbb{Z}
%EndExpansion
_{2}$ TSF has even fermion parity at $\mathbf{k}=(0,0)$ and $\mathbf{k}%
=(\pi,\pi),$ odd fermion parity at $\mathbf{k}=(0,\pi)$ and $\mathbf{k}%
=(\pi,0)$. This $%
%TCIMACRO{\U{2124} }%
%BeginExpansion
\mathbb{Z}
%EndExpansion
_{2}$ TSF supports the robust Majorana edge modes and there exists a new type
of low energy excitation - (supersymmetric) $%
%TCIMACRO{\U{2124} }%
%BeginExpansion
\mathbb{Z}
%EndExpansion
_{2}$ link-excitation. In the end we address a possible realization of such
interacting p-band fermions - a mixture of ultracold bosons on s-band and
spinless fermions on p-band in 2D square optical lattice.

The authors would like to thank W. M. Liu, C. Wu, H. Zhai, S. Chen for helpful
conversations. This work is supported by NFSC Grant No. 10874017, 11174035,
National Basic Research Program of China (973 Program) under the grant No. 2011CB921803.


\begin{thebibliography}{99}
\bibitem {note}The effective model of this Bose-Fermion mixture for the case
of $U_{s}>U_{\mathrm{sp}}>0$ is lightly different from that in Eq.[1] due to
the band-depedent anisotropic nearest neighbor attracting interactions.
However, in mean field level, the two models have the similar global phase diagram
\end{thebibliography}
\end{document}